# Collection of Propagating Electromagnetic Fields by Uncoated Probe

Farbod Shafiei [*], Michael C. Downer

Department of Physics, The University of Texas at Austin, Austin TX 78712 USA

*Correspondence to: farbod@physics.utexas.edu

**Abstract:**

**Understanding light-matter interaction at the nanoscale by observation of fine details of electromagnetic fields is achieved by bringing nanoscale probes into the nearfield of light sources, capturing information that is lost in the far field. Although metal coated probes are often used for nearfield microscopy, they strongly perturb the electromagnetic fields under study. Here, through experiment and simulation, we detail light collection by uncoated fiber probes, which minimize such perturbation. Second-harmonic light from intensely-irradiated sub-wavelength sub-surface features was imaged to avoid otherwise dominating fundamental light background, yielding clear nearfield details through a 50 nm aperture uncoated probe with ~23 nm optical resolution. Simulations shows how a metallic coating distorts optical nearfields and limits optical coupling into the probe in comparison to an uncoated probe.**

Introduction:

As silicon-based electronics started facing physical challenges such as limited bandwidth, power dissipation and cross talk among carrier channels, photons and optics have been studied as alternative information pipelines and logic gates [1-4]. The nanoscale details of electromagnetic fields during the propagation and interaction became vital for the future of nano-optics, nano-optoelectronic devices [5-10], and quantum optics and computation with subwavelength interacting elements. [11, 12] Even if light could be used only as a transmitting element, the interaction with the electronics nanocircuit

input/output demands nanoscale understanding of electromagnetic field interactions. [5, 13]

Understanding and controlling electromagnetic fields at the nanoscale has been a driving force for e.g. optical information processing at the nanometer scale by metamaterials nanocircuits [14], electrical control of light by graphene plasmons [15] and optical analogue computation by nanophotonics. [16]

Historically diffraction-limited optical microscopy has been the foundation of imaging, characterization and diagnosis in diverse fields such as living cells, materials science and microstructures. Optical microscopy started from observing microorganisms in 1665 and reached nano-resolution in recent decades by overcoming the diffraction limit. [17, 18]

Nearfield probe microscopy and farfield super resolution microscopy have been the frontier of sub-diffraction-limited imaging with many breakthroughs. These include single molecule detection with ~12 nm resolution by nearfield scanning optical microscope (NSOM) with a coated aperture probe [19], imaging intracellular fluorescent proteins by photoactivated localization microscopy (PALM) [20] with ~20 nm resolution, and fluorescent imaging of proteins by stimulated emission depletion (STED) microscopy with ~20nm resolution. [21]

Farfield super resolution microscopy techniques such as PALM and STED are limited to biological samples. Moreover their imaging method does not involve direct interaction with electromagnetic fields at the nanoscale. Nearfield probe microscopy, on the other hand, brings a nanoscale probe into direct contact with nanoscale details of the fields, in principle enabling extreme resolution. Probe microscopy has mostly been implemented by transmitting light through a metal-coated probe, by minimizing the area of excitation near the coated probe's aperture or by collecting non-propagating (evanescent) waves at the surface of the sample with a coated probe. This approach comes with the disadvantage that the metal coating significantly perturbs the studied fields. [22]. Alternatively, the uncoated fiber probe with tens of nm aperture has shown the capability of tens of nm optical resolving power. [23, 24]. For minimizing the metal coated probe's perturbation, approaches such as cloaking have been proposed. [25, 26]

Previous simulations of uncoated probes have focused almost entirely on the question of leakage of light propagating within the fiber and out through probe aperture/tip and tapered sides [27, 28]. Here we focus instead on the inverse question of light collection into the uncoated probe's 50 nm aperture, its coupling into a propagating fiber mode, and the electric field perturbation that accompanies these processes. Experimental data, supported by simulations, show optical resolution of ~23 nm and low light leak from side tapered area for the uncoated probe. Detailed simulation shows how an uncoated fiber probe has minimal effect on the electric fields while metallic coated fiber probe strongly perturbs the light electric fields around the probe with a poor feed through the coated probe.

Second harmonic probe microscopy (SHPM) was used to collect nanoscale optical signatures of samples. The experimental setup used near-infrared femtosecond laser pulses to excite the sample, and an uncoated silica scanning probe with 50 nm aperture to collect near-ultraviolet second-harmonic generation (SHG) light. Fundamental light scattered from subsurface dislocation defects then localized and frequency-doubled within the scattering medium (see the Method Section for details). By collecting SHG light, the setup filtered out strong reflected fundamental light, which otherwise would have dominated the detected signal [24].

We define aperture as the near flat fiber tip end with ~50 nm diameter created by melting, pulling and tapering process that serve as the dominant conduits for incoming light. See the second part of Method Section for more details about 50 nm aperture fiber probe and its melting/pulling process.

**Results and Discussion:**

We have previously demonstrated the fidelity and resolution of SHPM imaging. [24] Here we focus on uncoated probe collection and electric field (EF) behavior in the presence of this probe. To support experimental observations, we carried out finite element simulations of the collection of EF by uncoated and Al coated fiber probes. Figure 1 shows simulated data for EF behavior with no probe (figure 1a), and in the presence of 50 nm aperture uncoated silica fiber probe (figure 1b) and 100 nm Al coated silica fiber probe

(figure 1c). All fields are averaged over 3 spatial dimensions and time. The uncoated fiber probe perturbed the propagating fields much more weakly than the metallic coated fiber probe. The propagating fields for Figure 1 plots are S polarized 500 nm wide. Figure 1b shows that optical leakage from the tapered side of the uncoated probe is small and that light collection is mostly through the aperture, not the tapered sides. Light transmission through the uncoated probe is also much stronger than for the coated probe. The EF just 20 nm and 100 nm inside the coated probe are 0.62 V/m and 0.18 V/m, respectively, while fields at the same spots right inside the uncoated probe are 0.94 V/m and 1.04 V/m. Figure 1i plots the EF strength along the horizontal light propagation axis for no probe, and along the centers of non-coated and coated silica probes, corresponding to plots a-c. These profiles show that light transmission is higher through uncoated fiber probes than through coated probes while coated probes strongly perturb the propagating electric fields. Back reflection from the coated probe (Figure 1 c) creates a high-contrast interference pattern in front of the probe, while the corresponding pattern is much weaker for the uncoated probe (figure 1b). At the same time the metallic coating helps keep the probe clear from side optical leakage. See supplementary data S-1 for more details.

To replicate different intensity of EF, we used propagating EF with different widths and checked the feed through of the probe by plotting the power-flow (W/m$^2$) streamlines. Figure 1d-f shows the 500 nm, 200 and 50 nm wide propagating fields coupling to the 50 nm uncoated probe. The power-flow streamlines show that the aperture area of the uncoated probe is the main coupling part even for weak electric fields (figure 1f), while the streamlines are flowing away from the probe around the taper side area. In contrast to a tapered probe, a cylindrical silica tube of 50 nm (figure 1 g and h) allows strong light leakage through its sides. Consequently, the probe's taper is critical in confining collected light and preventing leakage (figure 1b).

Figure 2 shows the results of experiments in which an uncoated 50 nm aperture probe scanned over GaAs thin films on Si, collecting SHG light generated by femtosecond laser irradiation. These films have crystallographic dislocation defects due to lattice mismatch at the interface of the grown GaAs crystal and the Si substrate. The defects act as scattering sites for 780 nm pulsed laser light. The elastic scattered light becomes localized, and the resulting interference [29] creates optical hotspots. [24] By collecting 390 nm signal from these hotspots, we avoid the otherwise dominating fundamental 780 nm laser light (see Method Section for details). Figure 2 a-c shows SHG (390 nm) signatures of these hotspots at different average laser power excitation (50 mW to 5 mW). The SHG hotspots maintain their shape and pattern for strong and weak signals, suggesting that collection is mostly through the probe aperture. This was supported by the simulations of probe coupling for a variety of intensities (figure 1d-f).

We raster scanned over 2x2 µm of the same sample with the laser focal spot centered on the probe axis and again when stationed ~5 µm to one side of it (figure 2 d-f). This was to check if light leaks through the tapered side of the probe or is mostly collected through its 50 nm entrance aperture. By moving the laser focal spot ~5 µm we could observe that most of the features and pattern maintain their original shape with small changes showing the collection of the SHG light has to be mostly through the 50 nm aperture. If there was a major leak from the large area of the taper side, moving the laser spot to the side should strongly affect the collected pattern and shape. An intensity drop is observed as the focal spot has weaker intensity at the side.

Supplementary data S2 shows a variety of probe shapes to check the leakage of propagating electric fields into the probe from the side area. A 50 nm wide fiber tube (S2 a) shows that simply having a narrow aperture does not keep the external field from leaking into the probe side (as figure 1 g, h). The wider 200 and 400 nm wide aperture fiber tube show that the wider probe cross section helps to push away the propagating fields away from the side of the fiber tube (s2 b, c) but such a probe does not have a high resolving power as a 50 nm tapered probe. Combining 50 and 400 nm fiber tubes (figure S2 d) does not help to prevent side leakage. A tapered shape on one side (figure S2 e) does show that propagating waves get pushed away from that side while there is more leak on the non-tapered side of the probe. The top tapered side collects a total 0.633 V/m electric field while the low non-tapered side collects 0.778 V/m for this configuration. Figure S2 f-i shows a series of taper shape probes with different cone angles. This series shows propagating EF get pushed away from the probe side with minimum leakage by optimum half cone angle of ~20 $^O$ of 0.565 V/m leakage on one taper side (S2 g). Beyond this taper angle, the leakage becomes stronger again. Figure S2 i shows the side leakage for a very low angle (~5 $^O$) is 1 V/m on one taper side. Probe leakage starts acting almost like a 50 nm flat fiber tube when half cone angle goes below ~3 °. The probe with 45°half cone angle (S2 f) has 0.672 V/m leakage on one taper side.

To check the precision of collection of the 50 nm aperture probe, we measured its optical resolving power. For this purpose, we used a known pattern with sharp edges made of alternating stripes of InP and SiO2. Here InP is a strong source of 390 nm SHG, SiO2 a negligibly weak source. Figure 3a shows a 2x2 µm SHPM scan of the InP-SiO2 sample with ~70 nm stripe width. The Scanning Electron Microscope (SEM) micrograph in Figure 3b shows detailed features of the stripe pattern. The sample surface was flattened through mechanical/chemical polishing, which was essential for avoiding any optical artifact due to topographical height differences. Figure 3c shows an SEM micrograph of the uncoated 50 nm aperture silica probe made through the melting and pulling process (see Methods Section for details). Figure 3d shows a profile cut of a typical sharp edge of the SHPM stripe pattern. The derivatives of two such edges (Figure 3e, f) yield point spread functions (PSFs), which in turn yield optical resolution of 23 and 27 nm. These resolving power showing the 50 nm aperture of the probe serve as the dominant conduits for light collection proving light leakage from the side of the probe is minimal. A finite element simulation replicated the observed stripe pattern and its sharp edges using weak sources of EF ~80 nm apart (Figure 3g). Calculation from the simulated edges shows ~29 nm wide PSF (Figure 3h, i), similar to the measured resolution.

To further compare uncoated and coated probes, we simulated two weak 30 nm wide sources of P-polarized EF separated by 80 nm while 50 nm uncoated and Al-coated probes scanned over them 20 nm above the sample plane. Figure 4a-d shows the results for the uncoated probe. The EF couples into the 50 nm probe aperture, but the uncoated probe perturbs its shape and intensity minimally. The two light sources are well resolved, and the signal profile approximates a simple convolution of the source and aperture shapes. Figure 4e-h shows results for the coated probe. Here the 100 nm Al coating strongly perturbs the EF. The shape and intensity of the EF changes to such an extent that in some cases the two separated sources of EF are not distinguished.

Figures 4i, j plot profile cuts of the EF 18 nm above the sample surface (2 nm below probe aperture) for the configurations shown in Figures 4b and f, respectively. For the metal coated probe (Fig. 4j), the 2 sources of EF appear as 4 separated EF peaks, while the intensity increases due to plasmonic effects. For the uncoated probe (Fig. 4i), the two 80 nm separated sources maintain their original pattern.

Supplementary data S-3 shows how the separated EF sources would be affected by the distance of the probe from the EF sources plane. The data show the approach of the uncoated (top row) and coated probe (bottom row) for 20 to 140 nm gap distance. The approach of coated probe strongly perturbs the EF sources shape in comparison to uncoated probe.

**Conclusion:**

We demonstrate that a second-harmonic probe microscope (SHPM) that collects SHG light from fs-laser-irradiated surfaces with a 50 nm uncoated scanning fiber probe tip images sharp-edged features with ~ 23 nm resolution. We observed in both experiments and simulations that 390 nm light couples into an uncoated appropriately-tapered dielectric probe mainly through its 50 nm aperture. Experimental data shows SHG features maintain their shape and pattern over a wide range of excitation intensity, despite lateral offsets of excitation and probe tip, showing that light collection occurs mostly though the 50 nm aperture, not the tapered sides. Simulations corroborate the conclusion that SHG light couples predominantly through the 50 nm tip, if it is appropriately tapered. They also show that an uncoated probe perturbs near-surface electric fields much less than an equivalently-shaped metal-coated probe, enabling more faithful scanned images of sub-wavelength features. Overall, the study shows that an uncoated tapered probe with 50 nm aperture is the right microscope tool to study the propagating electromagnetic waves with resolution below the diffraction limit.

**Method Section:**

1) System: SHG optical study of semiconductor thin film was performed by a fiber based nonlinear nearfield scanning optical microscope (NSOM) system with 50 nm fiber aperture. This uncoated probe scanning approach was used to avoid and minimize any enhancement and perturbation of the electromagnetic field at the probe area. The probe was kept at ~ 20-30 nm above the sample with a feedback loop system monitoring the amplitude of the scanning probe. A 76 MHz laser with ~150fs pulse width at ~780nm was focused on an area about ~10 μm on the sample and collection was done at 390 nm. The incident excitation angle is ~45° with P polarization. The Sample was scanned by a piezoelectric stage under the stationary probe. Fiber probe which supports light only below 600 nm picks up the propagating SHG signal of III-V film at nearfield regime and the signal gets filtered for residue of fundamental light. Photomultiplier tube (PMT) sensitive only to photons at 200-700 nm range was used with a photon counter system to measure the photons. The nonlinear response of the GaAs-Si film is typically ~$10^{15}$ times weaker than the fundamental reflection of the film. Nonlinear scanning probe microscope systems have advantages of having very high spatial resolution and being noise free by collecting the SHG signal away from the fundamental signal. Studying in the fundamental regime with excitation at 780 nm has a dis-advantage of not being able to distinguish the very weak scattered and localized light which create those hotspots. If we use the probe microscopy to look at the same wavelength of excitation light, the reflection of the excitation light at the surface dominates all the intensity information and the weak scattered and localized light at sub-surface dislocation area would not be distinguished. Instead of the linear study, if we look at the nonlinear response of the III-V film at 390nm, then there would not be such a problem of dominating reflection light from the surface of the sample as the only reflected and dominating signal at the surface is 780 nm which is filtered. Then scattered and localized SHG light intensity can be distinguished from the film typical SHG background response. This filtering approach is capable of distinguishing the very weak SHG scattered and localized hotspot in the presence of dominating surface reflection. The high resolution of the scanning probe microscope was explored and confirmed. [25]

2) Probe with 50 nm aperture: ~ 50 nm aperture of silica probe was created by melting, pulling, and tapering process (electron micrograph - Figure 2c). The melting/pulling and tapering process was done by a Sutter P-2000 micropipette puller machine. These parameters were used for optimum results: Heat 340, Filament 0 Velocity 15 Delay 128 Pull 150 with the pulling process time of 0.13 second. The fiber has 125 μm cladding diameter and 2 μm core diameter. The numerical aperture of core is 0.13

Data in Figure 2 shows that such near-flat regions of uncoated fiber tips serve as the dominant conduits for incoming light, enabling resolution of tens of nm. The SHG probe microscopy scans of control samples possessing a well-defined stripe pattern (sharp edge) demonstrate a spatial resolution of ~23 nm.

Prior experimental work [23, 24] showed that such near-flat regions of uncoated fiber tips serve as the dominant conduits for incoming light, enabling resolution of tens of nm.

**Acknowledgments:**

The work was supported in part by the Welch Foundation grant F-1038. The authors acknowledge the III-V thin film samples preparation and diagnosis by SEMATECH.

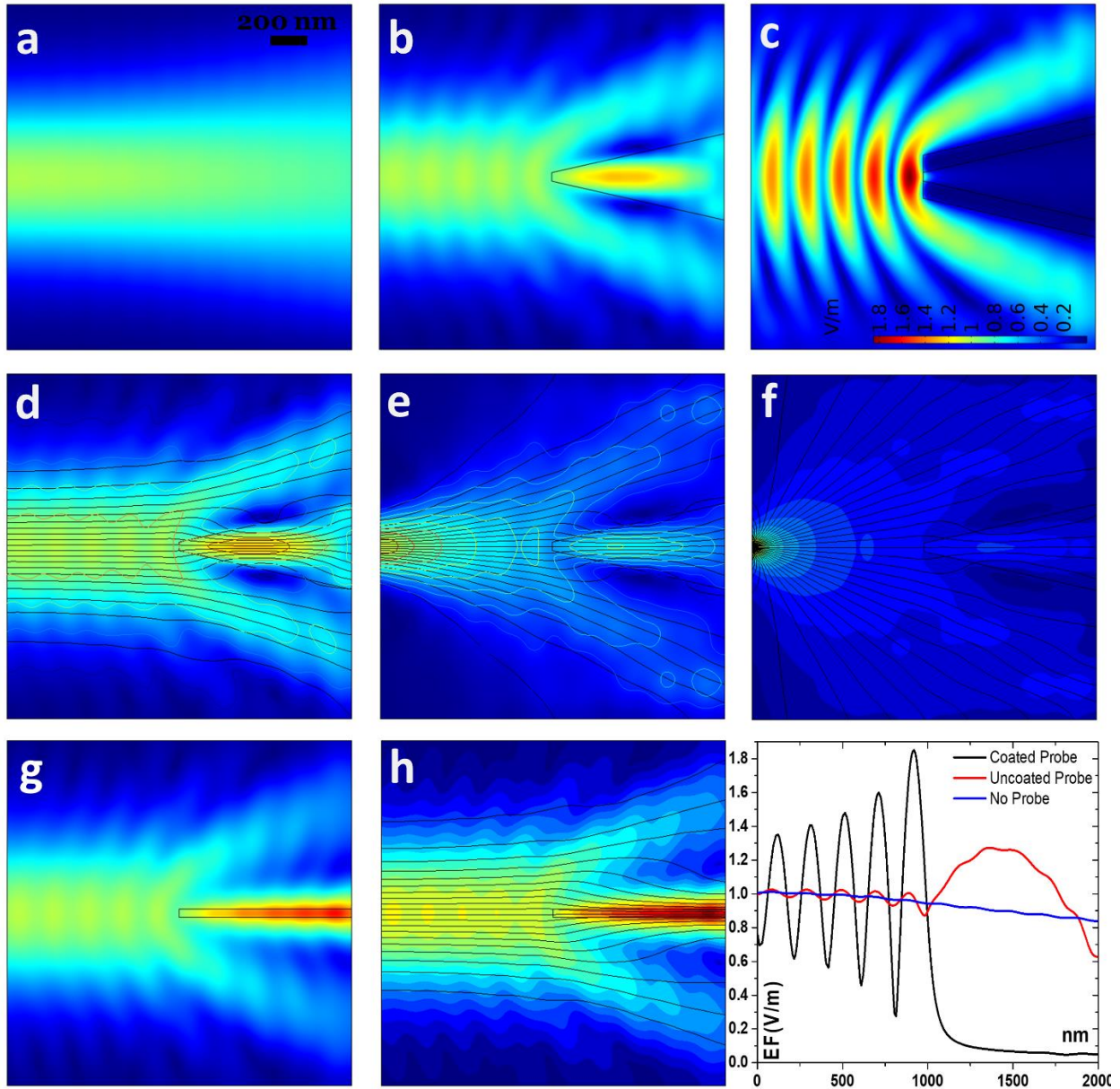

**Figure 1) Simulation of perturbation of propagating electric fields by probe.** No probe, unperturbed propagating EF (at 390 nm) **(a)**. Uncoated 50 nm aperture silica probe with weak perturbed propagating EF and weak leakage from taper sides and strong feedthrough **(b)**. 100 nm Al coated silica probe with strongly perturbed propagating fields and weak feedthrough **(c)**. Simulated EF and power-flow (black lines) for 500nm, 200 nm and 50 nm wide propagating EF **(d-f)**. The light coupling is mostly through the 50 nm aperture for weak and strong fields. 50 nm fiber tube shows strong side optical leakage **(g,h)**. EF profile cuts through the center of a-c plots show the weak and strong perturbation of propagating fields in front of the uncoated (red line) and coated probe (black line) **(i)**.

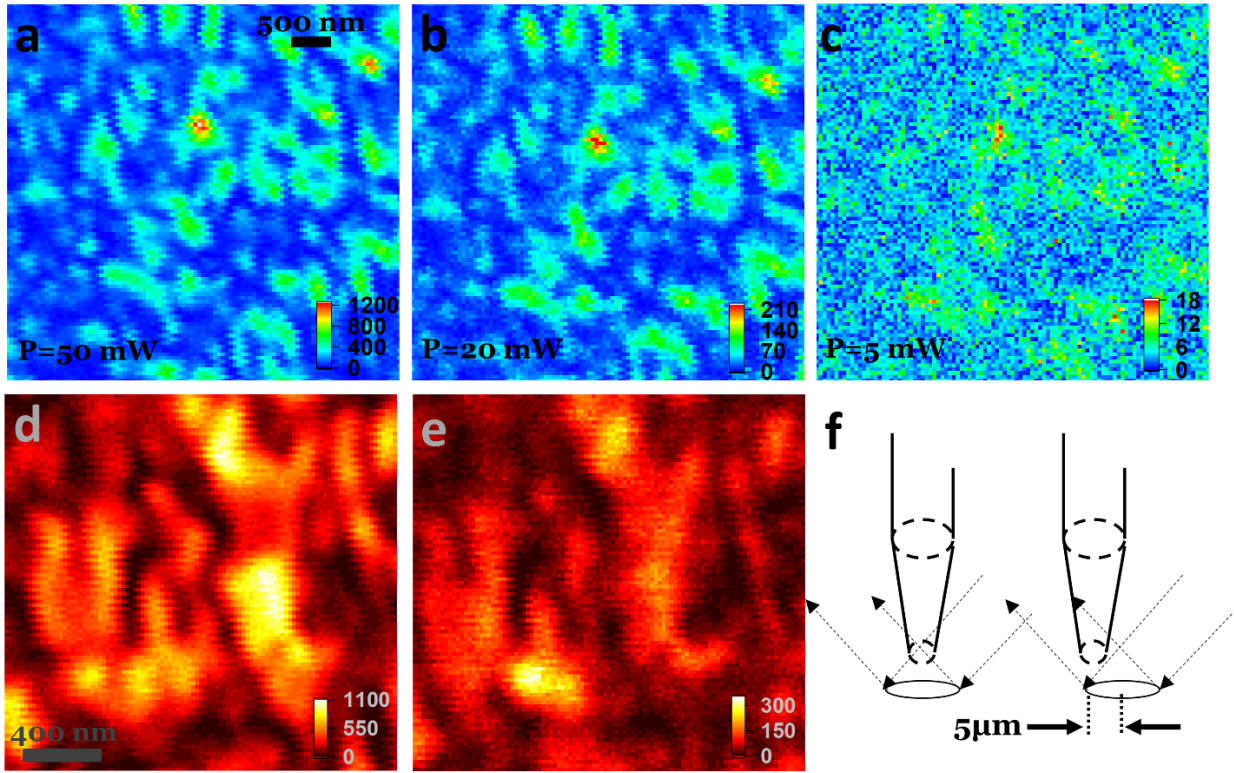

**Figure 2) Check for probe side leak during the collection process by different excitation power and steered focal spot.** 5x5 μm SHG Raster scan of GaAs-Si at different excitation power of 50, 20 and 5 mW by uncoated 50 nm aperture probe at ~ 20-30 nm probe-sample distance **(a-c)**. 2x2 μm SHG raster of the same sample while the laser focal spot was centered **(d)** and stationed ~ 5 μm to the side **(e)** with respect to probe axis (relative size of drawn parts are not to the scale) **(f)**. Shape and pattern of hotspots are maintained during these processes showing collection is mostly by 50 nm aperture area and light leak through the probe from taper sides is minimal.

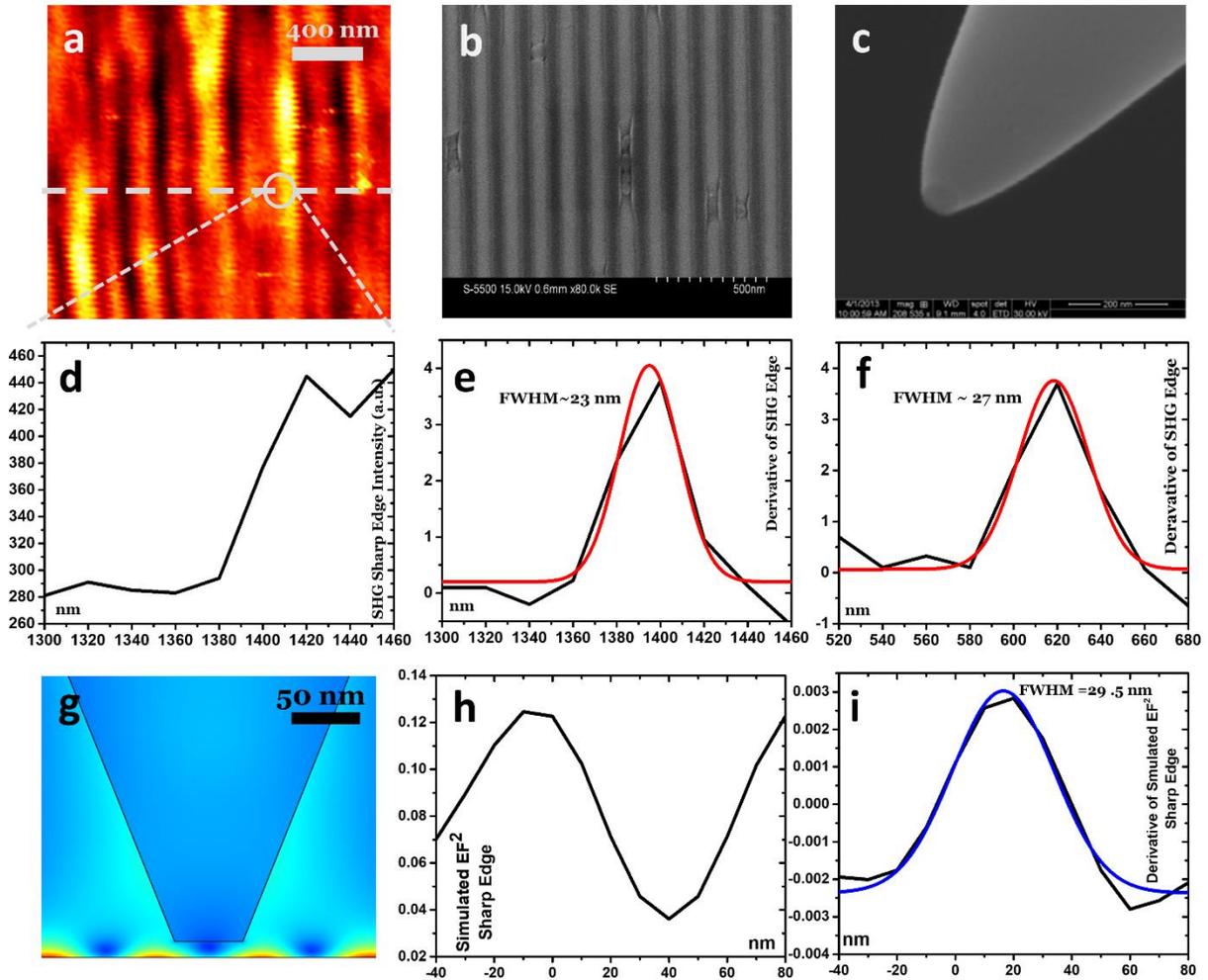

**Figure 3) Uncoated probe optical resolution.** Plot of SHG intensity as uncoated silica probe scans across stripe sample containing strong and weak sources of SHG light **(a)** SEM micrograph of striped InP-SiO2 sample **(b).** SEM micrograph of 50 nm silica uncoated probe **(c).** A profile cut **(d)** of the optical stripe pattern (a), showing a sharp optical edge which provides the PSF determining optical resolving power of 23 and 27 nm **(e,f).** Simulated EF collection of sharp edges **(g)**, replicating the collection of light through a 50 nm silica probe, provides us sharp edge and its simulated optical resolving power of 29 nm **(h,i).**

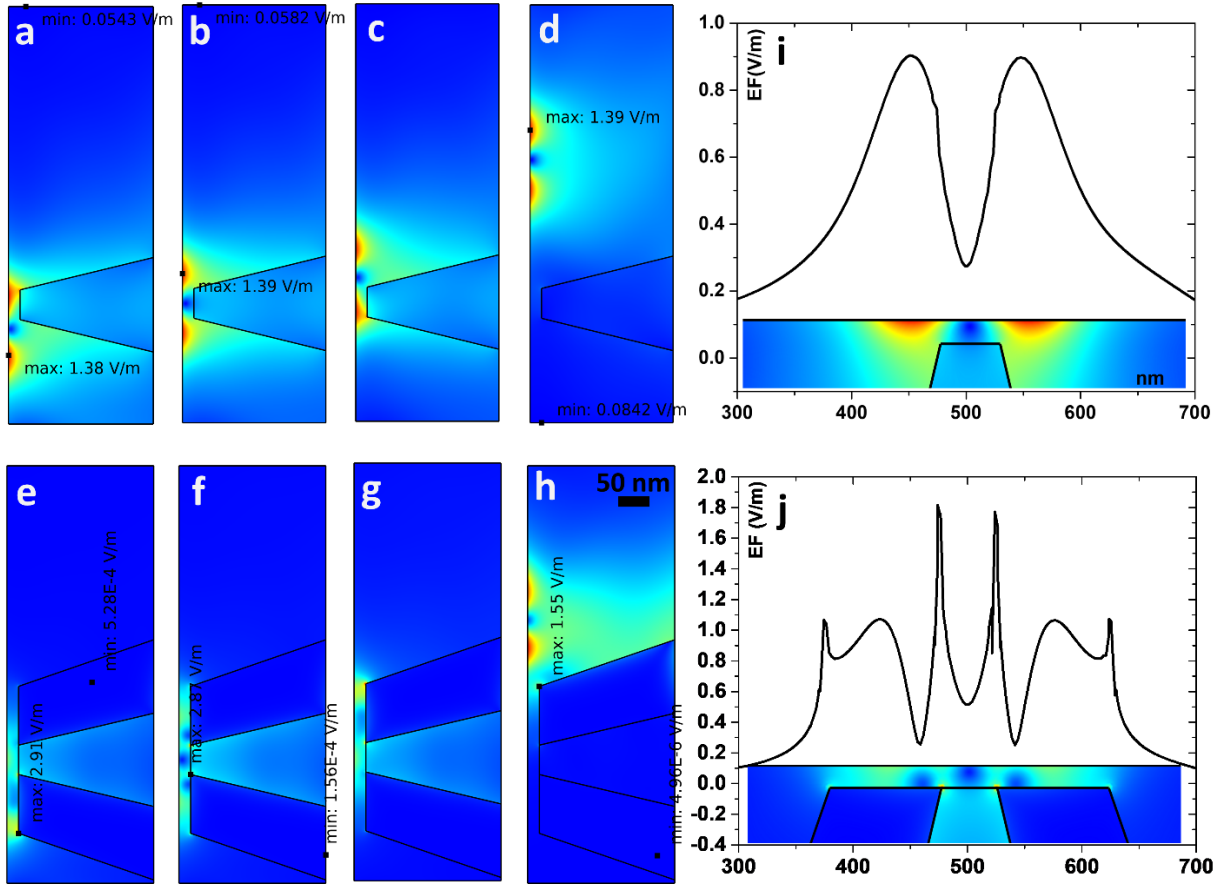

**Figure 4) Simulation of perturbation of electric field by uncoated and coated probes.** Simulation EF for uncoated 50 nm probe scanning over two electric field sources 80 nm apart **(a-d)** shows small perturbation of the field by the probe. Simulation of EF for 100 nm Al coated probe scanning over the same sources **(e-h),** shows stronger perturbation of the field. Profile cuts of electric field 18 nm above sample (2 nm below probe) for uncoated **(i)** and coated **(j)** probes.

# Supplementary Data for:
# Collection of Propagating Electromagnetic Fields by Uncoated Probe

Farbod Shafiei [*], Michael C. Downer

Department of Physics, The University of Texas at Austin, Austin TX 78712 USA

*Correspondence to: farbod@physics.utexas.edu

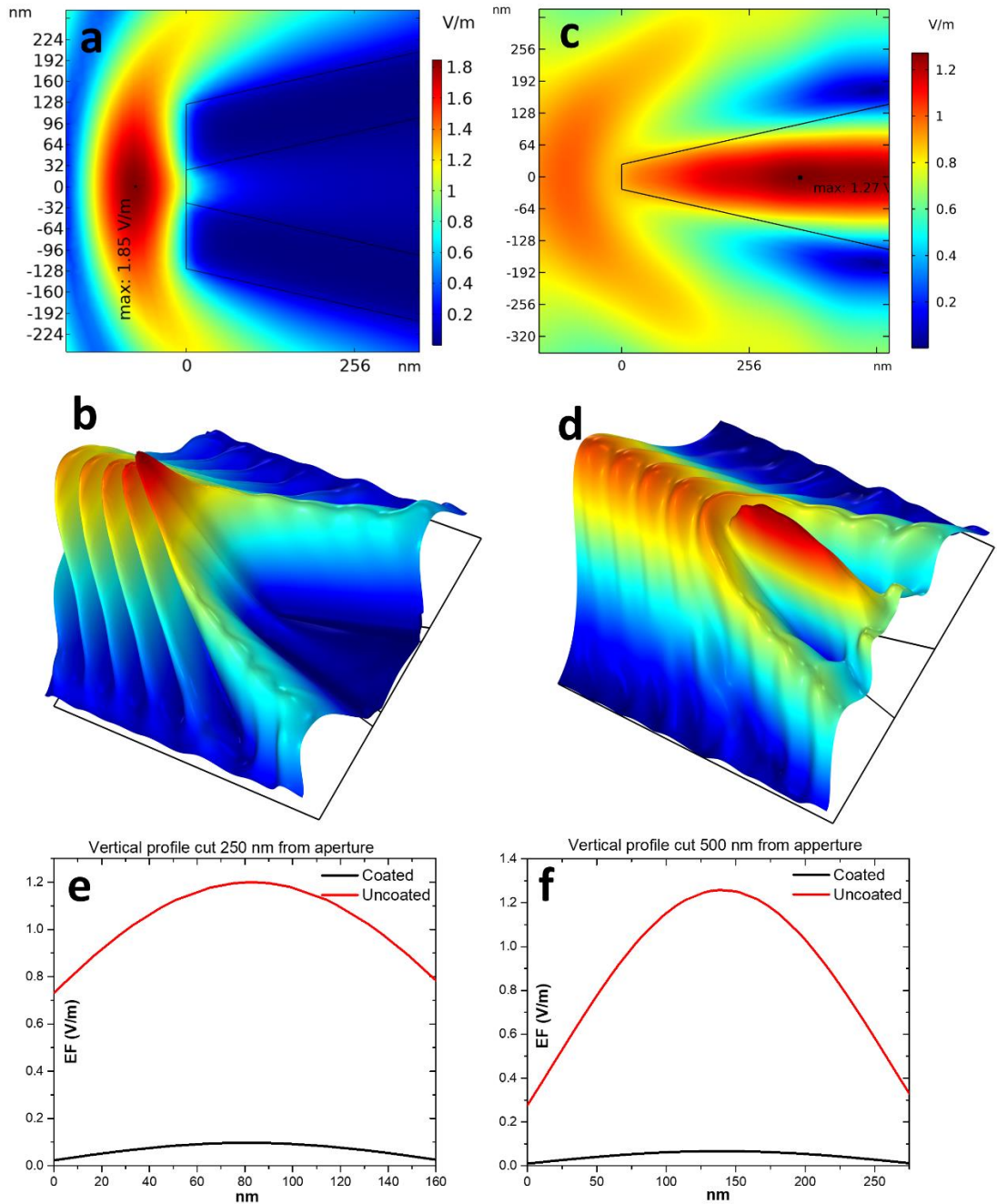

**Figure s1) Details of simulated EF in aperture/entrance area for coated and uncoated probe.** Coupling of propagating electric field with 100 nm Al coated silica probe with 50 nm aperture by finite element simulation **(a.b).** Coupling of propagating electric field with uncoated 50 nm aperture probe by finite element simulation **(c,d).** The 3D plots (b,d) show the strong perturbation of the coming field by coated probe while the light coupling is very weak for this probe. Vertical plots **(e,f)** are profile cut 250 nm and 500 nm away from the aperture for the coated and uncoated probes only inside the silica probe show intensity coupling is ~ 20 times weaker in the coated probe.

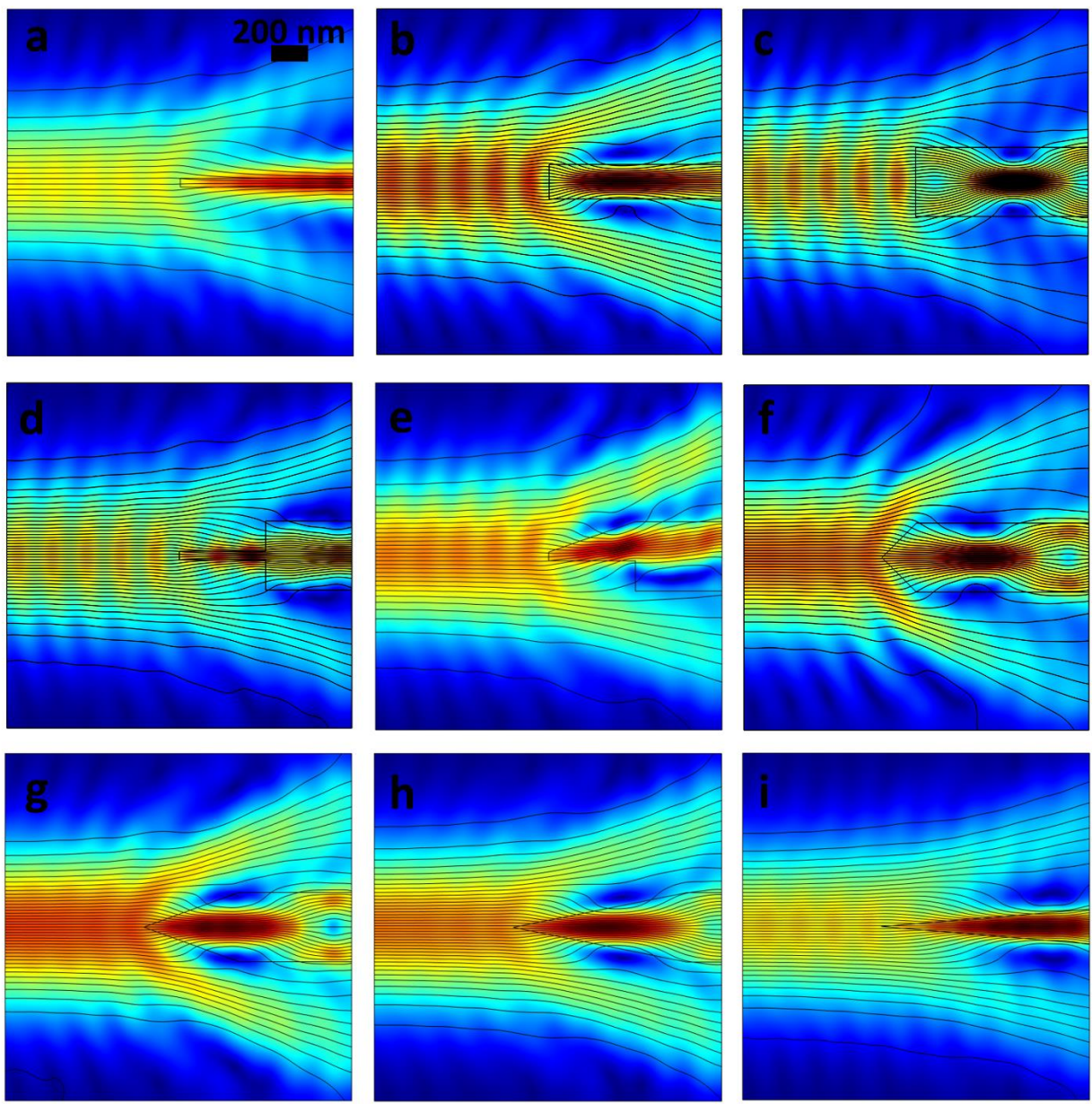

**Figure S2) Simulation of coupling of propagating electrified with different silica probes.** 50 nm fiber tube shows strong side optical leakage through finite element simulation **(a)**. Wider apertures of 200 and 400nm fiber tubes **(b, c)** show that such wide apertures push away the propagating waves and protect the probes from side leakage. Combining the 50 nm tube and 400 nm tube **(d)** does not help prevent side leakage of light. Tapering the top part of this combined probe **(e)** does collect less (0.633 V/m) electric field compared to the non-tapered lower side (0.778 V/m). Different probe cone angle **(f-i)** shows optimum half cone angle is ~ 20 º (g) with minimized light side leakage. Study of SEM graph (Fig.3c) shows the half cone angle of ~18-20 for the pulled fiber probes.

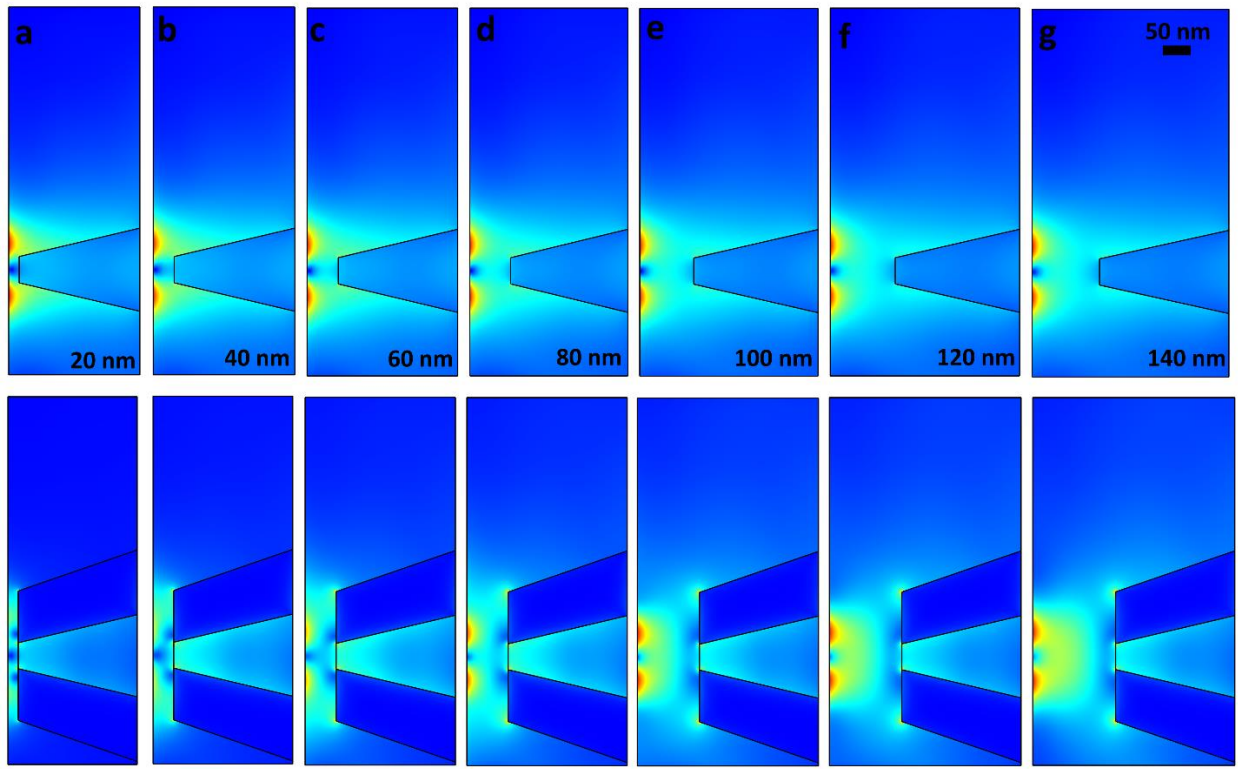

**Figure S3) Simulation of manipulation of EF field by coated and uncoated probe as function of height.** Series of height study **(a-g)** shows the EF perturbation of two separated sources by uncoated probe (top row) and 100 nm Al coated probe (bottom row). plots (a-c) for coated probe show this perturbation clearly. The presence of the uncoated probe 20 nm from surface (a) shows minimal perturbation compared to when the probe is at 140 nm distance (g) while the coated probe presence shows a distinct perturbation. The 80 nm separated two sources of electric field are unrecognizable when coated probe is very close to the source's surface (a).